\documentclass[]{article}
\usepackage{amscd,amsmath,amssymb,emlines2}
\title{COMPLEMENTARITY OF KINEMATICS AND GEOMETRY IN GENERAL RELATIVITY THEORY}
\author{Sergey S. Kokarev\thanks{logos-center@mail.ru}}
\date{RSEC "Logos",\, Central Post Office, cell 169, Yaroslavl, 150000, Russia}

\begin{document}

\maketitle

\begin{abstract}
Relations between kinematics, geometry and law of reference frame
motion are considered. We show, that kinematical tensors define
geometry up to a space functional arbitrariness when integrability
condition for spin tensor is satisfied. Some aspects of
geometrization principle and geometrical conventionalism of
Poincar\'{e} are discussed in a light of the obtained results.
\end{abstract}

\bigskip


\section{Introduction}

According to the equivalence principle which lies at the core of
the General Relativity (GR), inertia forces and gravity forces
cannot be distinguished from one another if one stays within the
boundaries of the theory. In general case semi-Riemannian metrics
of space-time describes a gravi-inertial complex which only in the
particular case of flat metrics (i.e. when the curvature tensor is
zero) can be interpreted as a pure field of inertia and as such
can be globally removed by choosing  the suitable reference frame.
In the case of space-time with a non-zero curvature gravitational
and inertial degrees of freedom are mixed and can be go over into
one another when  the reference frame is changed.

Classical approach to reference frames in the relativity theory
\cite{1,2} means to specify a world line of the referent body  (a
single observer) or local space-time bundle field of 1-form
(monad's formalism) or a tetrad of 1-forms (tetrad's formalism) on
a manifold with a specified metric $g$. The kinematic
characteristics of the reference frame (the world line curvature
for a single observer or forms of acceleration, rotation and
strain together with their components for monad's and tetrad's
formalisms) are calculated according to the standard formulae that
explicitly or implicitly include metric of the manifold. In
practice the tasks of defining (measuring) of metrics and the
kinematic characteristics of the reference frame go together.
Metric measurements (for example, by light signals in
chronogeometry, \cite{sing}) are based on certain hypotheses of a
concrete reference frame in which the measuring is taking place,
while the kinematic characteristics of that same reference frame
are to be defined according to the hypotheses of geometry.
Consequently there arises the question of the correlation between
kinematics and geometry within the GR, as well as that whether
they can be seen as interdependent  and maybe even
interchangeable. Let us explain the point with a simple example.
We'll consider a world line $\gamma$ of a test particle with
4-velocity field $u.$ According to the well known theorem of
geometry of manifolds \cite{arno1}, there can be found a
coordinate system, such as
\begin{equation}\label{str1}
u=\left.\frac{\partial}{\partial t}\right|_{\gamma},
\end{equation}
and, considering the normalization condition of the vector $u$:
$g(u,u)=1,$  the component $g_{tt}=g_{00}=1$ in the new coordinate
system. The question whether the test particle is going to move
with or without acceleration, if the metric $g$ is specified, can
be then easily solved. One needs to calculate the first curvature
of the $\gamma$ curve according to the formula:
\[
a=\nabla_uu,
\]
where the covariant derivative $\nabla$ is concordant with the
metric $g$. Of course, the result of the calculations doesn't
depend on the coordinate system chosen. Let us now consider a more
general problem definition, where we can change the geometry
(assuming, for example, it to be unknown and to be defined
experimentally), while the coordinate system remains constant (if
we assume that it is related to a reference frame of fixed
bodies). Let us consider the metric $g',$ which in the coordinate
system that straighten   the field $u$ (i.e. where (\ref{str1}) is
true) takes the form:
\begin{equation}\label{metr1}
g'=dt\otimes dt-h_{ik}dx^i\otimes dx^k\quad (i,k=1,2,3).
\end{equation}
According to the well known theorem of differential geometry (for
example, see \cite{oneil}), on the manifold with such metric the
line $\gamma$ will be a geodesic. Therefore, its curvature which
is equal to 4-acceleration turns to zero. So, through a choice of
a suitable geometry such important kinematic characteristic of
motion as particle acceleration can be turned to zero (or, if we
prefer a more general definition, can be made equal to any
4-vector, specified beforehand and orthogonal to $u$ --- see
below). Basically this possibility expresses the essence of the
geometrization principle of the physical interactions, which is
the ground principle in modern theories, considering physical
interactions.

These conclusions can be easily applied to the case of 4-velocity
field which describes continuous medium, that specifies an
extended reference frame. However, in the case of the continuous
medium the analysis is complicated by the fact that besides the
curvature vector of congruence, such systems have a couple of
additional characteristics as well, such as congruence rotation
tensor (or spin tensor) $\omega$ and rate of deformation
$\mathcal{D}$ tensor, which impact the calculations when applied
in practice.

Therefore if we reject a priori assumptions related to background
geometry and kinematic characteristics of the reference frame,
there appears a construction which consists of three "mobile"\,
parts: geometry (manifold metrics), reference frame (time lines
congruence) and kinematics (kinematic tensors). In this case the
standard definition of the problem considering determination of
kinematic tensors related to decomposition of a covariant
derivative of the field of the reference frame into spacetime
projections ($\tau$-field, see formula (\ref{decf}) below) is only
one of the three fundamental problems shown on the diagram
\ref{dg1} below.

\begin{figure}[htb]
\centering \unitlength=0.50mm \special{em:linewidth 0.4pt}
\linethickness{0.4pt} \footnotesize \unitlength=0.70mm
\special{em:linewidth 0.4pt} \linethickness{0.4pt}
\unitlength=1mm \special{em:linewidth 0.4pt} \linethickness{0.4pt}
\begin{picture}(66.83,36.00)
\emline{0.67}{12.50}{1}{0.67}{1.17}{2}
\emline{0.67}{1.17}{3}{25.83}{1.17}{4}
\emline{25.83}{1.17}{5}{25.83}{12.33}{6}
\emline{25.83}{12.33}{7}{0.67}{12.33}{8}
\emline{42.17}{12.33}{9}{42.17}{1.17}{10}
\emline{42.17}{1.17}{11}{66.83}{1.17}{12}
\emline{66.83}{1.17}{13}{66.83}{12.17}{14}
\emline{66.83}{12.17}{15}{42.17}{12.17}{16}
\emline{21.50}{36.00}{17}{21.50}{23.17}{18}
\emline{21.50}{23.17}{19}{46.17}{23.17}{20}
\emline{46.17}{23.17}{21}{46.17}{36.00}{22}
\emline{46.17}{36.00}{23}{21.50}{36.00}{24}
\emline{25.83}{6.83}{25}{42.17}{6.83}{26}
\emline{11.17}{12.33}{27}{34.00}{23.17}{28}
\emline{34.00}{23.17}{29}{55.33}{12.17}{30}
\emline{44.00}{18.00}{31}{25.83}{12.33}{32}
\emline{42.00}{12.17}{33}{23.33}{18.17}{34}
\emline{33.83}{23.17}{35}{33.83}{6.83}{36}
\emline{33.33}{19.33}{37}{33.83}{23.33}{38}
\emline{33.83}{23.33}{39}{34.33}{19.33}{40}
\emline{38.83}{13.83}{41}{42.17}{12.17}{42}
\emline{42.17}{12.17}{43}{38.33}{12.67}{44}
\emline{28.17}{13.67}{45}{25.83}{12.33}{46}
\emline{25.83}{12.33}{47}{29.17}{12.67}{48}
\put(44.00,18.00){\circle*{1.33}}
\put(23.50,18.17){\circle*{1.33}} \put(33.83,6.83){\circle*{1.33}}
\put(40.67,13.67){\makebox(0,0)[lb]{$1$}}
\put(26.33,13.67){\makebox(0,0)[rb]{$2$}}
\put(34.67,21.00){\makebox(0,0)[lc]{$3$}}
\put(33.83,34.17){\makebox(0,0)[ct]{Frame}}
\put(33.83,24.50){\makebox(0,0)[cb]{$(\tau$-field)}}
\put(13.00,6.83){\makebox(0,0)[cc]{Metric $g$}}
\put(54.67,11.00){\makebox(0,0)[ct]{Kinematics}}
\put(54.67,1.83){\makebox(0,0)[cb]{(tensors $a,$ $\omega,$ $D$)}}
\end{picture}
\caption{\small  Statements of the problems that explain relations
between geometry, reference frame and kinematics}\label{dg1}
\end{figure}
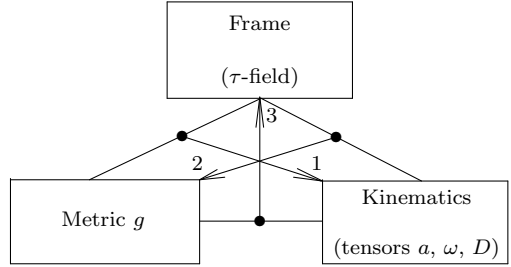

Let us sum up the statements of the problems as shown on the
diagram.
\begin{enumerate}
\item {\bf  Traditional (geometric) statement of the problem. }
Kinematic characteristics of the reference frame are being
calculated with regard to the specified metric $g$ and specified
congruence of the $\tau$-field. This problem has only one solution
on provision that the manifold metric is known a priori.
\item
{\bf  Kinematic statement of the problem. } After this statement,
it is the manifold metric that is being calculated on the basis of
specified congruence of $\tau$-field and a set of kinematic
tensors. As will be shown below, this problem always allows a
solution, provided the metric and kinematic tensors satisfy
consistency constraints (not overly rigid). Metric in this case
has more than one definition.
\item
{\bf  Mixed (kinematic-geometric) definition of the problem.} Here
we reconstruct $\tau$-field after specified metric $g$ and
specified kinematic tensors. In order to obtain a solution we need
that the metric and kinematic tensors satisfy rather rigid
conditions (integrability conditions of kinematic tensors).
\end{enumerate}

All three statements of the problem have a fundamental value and
practical applications. The first, geometric statement arises when
one calculates the observable effects of the GR, based on the
known metrics (for example, with a certain type of symmetry and
asymptotic properties). The second (kinematic) statement can be of
use when one tries to define the geometry of space-time with the
help of experiments that are carried out in a certain reference
frame with known properties (see discussion in Conclusion).
Lastly, the third (mixed) statement is related to the definition
of the motion of the reference frame by the inertia fields
"measurements", based on the known geometry of space-time (see
discussion in Conclusion).

The aim of the present paper is to investigate two of the
unconventional definitions mentioned above, both of which bring to
light  the interdependence that exist between geometry and
kinematics in GR. Our analysis shows that there is a subtle
interrelation between geometry, kinematics and practice, and
mostly confirms the viewpoint that concerns the conventionality of
geometry suggested by A. Poincar\'{e} \cite{3}. Physical
implications of the results obtained, as well as some of the
principal conclusions concerning construction of different
physical theories in general will be discussed in Conclusion.

When relaying general statements of fundamental character we will
use free of coordinate formulations, mathematical notations of
which basically agree with those found in the well-known
guidebooks on differential geometry of manifolds
\cite{oneil,warner,griffits}. Particularly, we use the following
notations and abbreviations:

$\imath_X$ and $\jmath_\omega$ --- 1-form and vector field dual to
vector field $X$ and 1-form $\omega$ respectively. For any vector
field $Y$  we have: $\imath_X(Y)\equiv\langle X,Y\rangle,$
$\langle\jmath_\omega,Y\rangle\equiv \omega(Y),$ where $\langle\
,\ \rangle$ --- Riemannian metric; sometimes we'll use shortened
notation $\overrightarrow{\omega}$ for $\jmath_\omega$;

$T^r_s(\mathcal{M})$ --- fibering of  $r$-contravariant and
$s$-covariant tensor fields over $\mathcal{M}$;

$\mathcal{T}(\mathcal{M})=\bigoplus\limits_{r,s}
T^r_s(\mathcal{M})$ --- tensor algebra over $\mathcal{M}$;

For any $T\in T^0_2(\mathcal{M})$ we define $\jmath T$ and
$T\jmath$ by the formulae:
\[
(\jmath T)(\omega, X)=T(\jmath_\omega,X);\quad
(T\jmath)(X,\omega)=T(X,\jmath_\omega).
\]
Coordinate form: $(\jmath
T)^\alpha_\beta=G^{\alpha\gamma}T_{\gamma\beta},$ $(T
\jmath)^\alpha_\beta=G^{\alpha\gamma}T_{\beta\gamma}$ shows, that
$\jmath$  can be viewed as coordinate free notation of tensor
indexes raising. Lowering is defined similarly by means of
$\imath;$

$\hat{\mathcal{S}}$ and $\hat{\mathcal{A}}$ --- symmetrization and
antisymmetrization operators, acting in spaces
$T^n_0(\mathcal{M})$ and $T_n^0\mathcal{M}$ for every $n$; for
example, in case $T\in T^0_2(\mathcal{M}):$
\[
(\hat{\mathcal{S}}T)(X,Y)=\frac{1}{2}(T(X,Y)+T(Y,X));\quad
(\hat{\mathcal{A}}T)(X,Y)=\frac{1}{2}(T(X,Y)-T(Y,X));
\]

$a\vee b\equiv a\otimes b+b\otimes a$ --- symmetrized tensor
product of vectors or 1-forms;

$\nabla:\ T^r_s(\mathcal{M})\to T^r_{s+1}(\mathcal{M})$ covariant
(with respect to some fixed Riemannian metrics $g$) derivative;

$\pi_X:\ \Lambda^p(\mathcal{M})\to\Lambda^{p-1}(\mathcal{M})$ ---
lowering degree operator, acting on space of external forms of
degree $p$ by the rule:
\[
(\pi_X\omega)(Y_1,\dots,Y_{p-1})=\omega(X,Y_1,\dots,Y_{p-1});
\]

$\Delta\equiv\det(g)$ --- determinant of the metric tensor matrix;

$\text{vol}\equiv\sqrt{|\Delta|}\,dx^0\wedge dx^1\wedge dx^2
\wedge dx^3$
---
standard volume form on 4-dimensional manifold,
\[
\widetilde{\text{vol}}\equiv\frac{1}{\sqrt{|\Delta}|}\,\partial_0\wedge\partial_1\wedge\partial_2
\wedge \partial_3
\]
--- invariant 4-vector ("volume form in space of 1-forms").

$\ast$:
$\Lambda_s(\mathcal{M})\leftrightarrow\Lambda^{4-s}(\mathcal{M})$
--- covariant operator of dualization isomorphism, mapping fibering of external forms of rank $s$ in fibering
polyvectors of rank  $4-s$ and vice versa.   For example, 3-form
$\ast X\equiv\pi_X\text{vol},$ and  2-form $\ast(X\wedge
Y)\equiv(\pi_Y\circ\pi_X)\text{vol}$ and is uniquely defined by
its values on the pair of vectors $Z,W$:
\[
\ast(X\wedge Y)(Z,W)\equiv\text{vol}(X,Y,Z,W);
\]

$\ell(a,b,c,\dots)\equiv\text{span}(a,b,c,\dots)$ --- span of
elements $a,b,c$ of some linear space.

We use Greek indices $\alpha,\beta,\gamma,\dots=0,1,2,3$ for the
space-time components of geometrical objects  and Latin indexes
$i,j,k,\dots=1,2,3$ for a purely  space components of geometrical
objects.

\section{Essentials of reference frames theory in GR}

The aim of present section is to expose monad's formalism of
general relativity (GR) with some details necessary for future
purposes.

\subsection{Algebra of monad's formalism}

Let  $\mathcal{M}$ be (semi-)Riemannian 4-dimensional manifold
with some fixed metric $g.$ The most general way to define 3D
submanifolds ({\it submanifolds of simultaneous events}),
"embedded"\, into $\mathcal{M},$  is to fix some smooth  {\it
1-form of time $\tau$}  ({\it $\tau$-field, monad field}) with the
normalization condition:
\begin{equation}\label{norma}
\tilde g(\tau,\tau)=1,
\end{equation}
where $\tilde g$ is standard metric in space of 1-forms, induced
by the Riemannian metric $g.$ This form induces decompositions of
tangent and cotangent spaces at every point $p\in\mathcal{M}$:
\begin{equation}\label{decomp}
T_p\mathcal{M}=(T_p)_h\mathcal{M}\oplus \ell_p(\jmath_\tau);\ \ \
T_p^\ast\mathcal{M}=(T_p^\ast)_h\mathcal{M}\oplus \ell_p(\tau),
\end{equation}
where {\it space-like tangent and cotangent subspaces} are defined
by the formulas:
\[
(T_p)_h\mathcal{M}\equiv\{v\in T_p\mathcal{M}\,|\, \tau(v)_p=0\}
\]
and
\[
(T_p^\ast)_h\mathcal{M}\equiv\{\lambda\in T^\ast_p\mathcal{M}\,|\, \lambda(\jmath_{\tau})_p=0\}
\]
respectively. Subspaces  $\ell_p(\jmath_{\tau})$ and
$\ell_p(\tau)$ we'll call {\it tangent and cotangent time
directions at the point $p$.} Let's note, that the set
\[
T_h\mathcal{M}\equiv\bigcup\limits_{p\in\mathcal{M}}(T_p)_h\mathcal{M}
\]
(or similarly  $T^\ast_h\mathcal{M}$) in general does'nt admit
local representation $R\times T(\mathcal{M}_h),$ where
$\mathcal{M}_h$ is some space 3D manifold, since the form $\tau$
can be {\it anholonomic (nonintegrable)}. In this situation we'll
refer to $\mathcal{M}_h$ as {\it anholonomic horisontal manifold}
\cite{griffits}, such that formally\footnote{In case of the, so
called, {\it complete nonintegrability,} Rashevski-Chow's  theorem
\cite{rash,chow} states, that $\mathcal{M}_h=\mathcal{M}$ i.e. any
two points of $\mathcal{M}$ can be joined by a some horisontal
curve $\gamma_h.$} $T(\mathcal{M}_h)\equiv T_h\mathcal{M}.$

Tensor continuations of (\ref{decomp}) give decomposition of a
whole tensor algebra $\mathcal{T}(\mathcal{M})$ on   $\tau-h$
components. Formally, let consider linear operator (affinnor
field):
\begin{equation}\label{aff}
\hat h\equiv \hat I-\tau\otimes \jmath_\tau\equiv\hat I-\hat\tau,
\end{equation}
mapping $T\mathcal{M}\to T\mathcal{M}$ and $T^\ast\mathcal{M}\to
T^\ast\mathcal{M}.$ Here $\hat I={\rm id}_{T\mathcal{M}}$ or $\hat
I={\rm id}_{T^\ast\mathcal{M}}.$ By the definition it follows,
that $\hat h(\hat h(X))=\hat h(X)$  and $\langle\hat
h(X),Y\rangle=0$ for every vector field $X$ and every vertical $Y$
(the same is true for 1-forms). So, $\hat h$ is projector:
$T\mathcal{M}\stackrel{\hat h}{\to} T_h\mathcal{M}$ or
$T^\ast\mathcal{M}\stackrel{\hat h}{\to} T^\ast_h\mathcal{M}.$
Writing $\hat I=\hat \tau+\hat h$ and taking its $n$-th tensor
degree, we have:
\begin{equation}\label{decompun}
\hat I^{\otimes n}\equiv{\rm id}_{T^r_{n-r}(\mathcal{M})}=(\hat\tau+\hat h)^{\otimes n}
=\sum\limits_{\varsigma}\hat\pi_\varsigma,
\end{equation}
where $\varsigma$ runs all binary sequences of symbols
$\{\tau,h\}$ of length $n,$ $\hat\pi_\varsigma$ --- projector on
$\varsigma-$th component of $T^r_{n-r}(\mathcal{M}).$ Acting by
initial  and final operators of (\ref{decompun}) on any tensor
field $T\in T^r_{n-r}(\mathcal{M})$, we have
\begin{equation}\label{decompt}
T=\sum\limits_{\varsigma}T_\varsigma,
\end{equation}
where  $T_\varsigma=\hat\pi_\varsigma(T)$ is  $\varsigma$-th
projection of  $T.$ In what follows we'll denote projections by
index-like symbols $\tau$ or $h$  when it will not lead to
ambiguousness. For example, any vector field can be decomposed as
follows:
%\begin{equation}\label{decompvec}
\begin{equation}\label{vectdec}
X=X_\tau \jmath_{\tau}+X_h,
\end{equation}
%\end{equation}
where $X_\tau\equiv\tau(X),$ $X_h\equiv\hat h(X).$

With using (\ref{decompt}) it is easy to get  decomposition of
$g$:
\begin{equation}\label{dec}
g=\tau\otimes\tau-h,
\end{equation}
where $h$ is {\it metric on (anholonomic) manifold}
$\mathcal{M}_h,$ defined by the rule:
\begin{equation}\label{hprod}
h(X,Y)=-g(\hat h(X),\hat h(Y))
\end{equation}
for any vector fields $X,Y.$ Relation (\ref{hprod}) means, that
\[
h(X,Y)=-g(X,Y)
\]
for every horizontal vector fields $X=X_h$ and $Y=Y_h$ and
\[
\ker\,h=\ell(\jmath_{\tau}).
\]
%С позиций физики  $\tau$ определяет гладкое семейство локальных пространственных
%сечений, которые "высекают"\, 3-мерный мир внутри заданного 3-мерного.

We define space-like volume forms $\text{vol}_h$ and
$\widetilde{\text{vol}}_h$ by the relations:
\begin{equation}\label{vol3}
\text{vol}_h\equiv\ast\overrightarrow{\tau};\quad
\widetilde{\text{vol}}_h\equiv\ast\tau.
\end{equation}
This pair of forms induces {\it operation of space dualization}
$\star,$ mapping space-like forms and polyvectors with summarized
rank 3 in each other.

So, the  $\tau$-field allows one to make local differentiation of
space and time projections of any geometrical objects without more
detailed differentiation of space projections onto space
components. The latter problem can be solved within the frame of
tetrad formalism, which is referred to complete methods of
description of reference frame \cite{1}.

\subsection{1+3-analysis on $\mathcal{M}$}\label{analysis}

By (\ref{norma}) it follows, that%\footnote{Here and below  we use
%abbreviated notation $D_\omega\equiv D_{\jmath_\omega}$ for any
%kind of derivative $D$ along vector field  $\jmath-$conjugated
%with some $1-$form $\omega.$}
$(\nabla_\tau\tau)_\tau=0.$ Let
define {\it space curvature 1-form of $\tau-$congruence}:
\begin{equation}\label{defcurve}
a\equiv\nabla_{\overrightarrow{\tau}}\tau.
\end{equation}
From the view point of kinematics this form defines {\it
acceleration field} of reference frame and characterizes a measure
of deflection of integral curves related to the vector field
$\overrightarrow{\tau}$ from geodesics (straightest), which
corresponds to a free fall. Obviously, the tensor $
\mathcal{H}\equiv\nabla\tau-\tau\otimes a $ is space-like. It can
be decomposed on  symmetric and antisymmetric components:
%\begin{equation}\label{decsymm}
$\mathcal{H}=\mathcal{D}+\mathcal{\omega},$
%\end{equation}
where
%\begin{equation}\label{def}
\begin{equation}\label{defexc}
\mathcal{D}\equiv\hat{\mathcal{S}}(\nabla\tau-\tau\otimes a)=
\frac{1}{2}(L_\tau g-\tau\vee a)
%\end{equation}
\end{equation}
--- {\it space tensor of external curvature of the manifold
$\mathcal{M}_h$}, having kinematical sense of  {\it velocity
deformations field,} related to the reference frame $\tau,$ and
%\begin{equation}\label{rot}
\begin{equation}\label{deftors}
\mathcal{\omega}\equiv\hat{\mathcal{A}}(\nabla\tau-\tau\otimes a)=
\frac{1}{2}(d\tau-\tau\wedge a)
\end{equation}
%\end{equation}
--- {\it  space rotation tensor of  $\tau-$congruence},
having kinematical sense of  {\it local spin} of the reference
frame. So, finally we obtain the expression
\begin{equation}\label{decf}
\nabla\tau=\tau\otimes a+\omega+\mathcal{D}.
\end{equation}
Acting in (\ref{decf}) by $\jmath$ from the right (with using
$[\nabla,\jmath]=0$), we obtain for the vector field
$\overrightarrow{\tau}=\jmath_\tau$:
\begin{equation}\label{decv}
\nabla
\overrightarrow{\tau}=\tau\otimes\overrightarrow{a}+\hat{\omega}+\hat{\mathcal{D}},
\end{equation}
where $\hat{\omega}=\omega\jmath,$ $\hat{D}=D\jmath.$

Following to \cite{1}, let define operators of {\it time-like} and
{\it space-like}  {\it derivatives}:
%\begin{equation}\label{der}
\[\dot T_h\equiv\frac{d}{d\tau}T_h\equiv(L_{\overrightarrow{\tau}} T_h)_h;\ \ \ {}^{(3)}\nabla T_h\equiv(\nabla_h T_h)_h,\]
%\end{equation}
where $T_h$ is arbitrary space tensor field. On scalar functions
by definition we have:
\[
\dot f= \overrightarrow{\tau}(f);\ \ \ {}^{(3)}\nabla f=(df)_h\equiv d_hf.
\]
With using (\ref{decompt}) the following identity for any vector
field $Z$ can be established:
\begin{equation}\label{deccov}
\nabla Z={}^{(3)}\nabla Z_h+Z_\tau\hat{\mathcal{H}}+
(\dot Z_\tau-Z_a)\tau\otimes \overrightarrow{\tau}+
\tau\otimes(Z_\tau \overrightarrow{a}+(\dot Z_h+\hat{\mathcal{H}}(Z_h,\ ))
\end{equation}
\[
+((dZ_\tau)_h-\mathcal{H}(\ ,Z_h))\otimes \overrightarrow{\tau},
\]
where $Z_a=a(Z).$ Acting on (\ref{deccov}) by $\imath$ from the
right, identifying $\imath_Z\equiv\lambda$ and using the relation
\[
\imath_{\dot Z_h}=\frac{d}{d\tau}\imath_{Z_h}-2\mathcal{D}(Z_h,\ ),
\]
we have for 1-forms:
\begin{equation}\label{deccovf}
\nabla\lambda={}^{(3)}\nabla \lambda_h+\lambda_\tau\mathcal{H}+
(\dot \lambda_\tau-\lambda_a)\tau\otimes\tau+
\tau\otimes(\lambda_\tau a+(\dot \lambda_h-\hat{\mathcal{H}}(\ ,\lambda_h))
\end{equation}
\[
+((d\lambda_\tau)_h-\hat{\mathcal{H}}(\ ,\lambda_h))\otimes \tau.
\]
Assuming in (\ref{deccovf}) $\lambda=\tau,$ $\lambda_\tau=1,$
$\lambda_h=\lambda_a=0$ we obtain (\ref{decf}).

Formulas (\ref{deccov})-(\ref{deccovf}) show, that any 4D tensor
expression, including covariant derivatives, can be rewritten in
terms of time-like and space-like derivatives. The following
useful identities are easy checked:
\begin{equation}\label{formull}
\dot\tau=a=L_{\tau}\tau;\ \ \ {}^{(3)}\nabla\tau\equiv(\nabla_h\tau)_h=\mathcal H;\ \ \ \dot h=2\mathcal{D}=\dot g  ;\ \ \ {}^{(3)}\nabla h=0.
\end{equation}
The latter expression suggests, that operator ${}^{(3)}\nabla$
should be treated as "covariant"\footnote{In fact,
${}^{(3)}\nabla$ possesses effective torsion, since direct
calculation gives:
$\text{Tors}_{{}^{(3)}\nabla}(X_h,Y_h)\equiv{}^{(3)}\nabla_{X_h}Y_h-{}^{(3)}\nabla_{Y_h}X_h-
[X_h,Y_h]=2\omega(X_h,Y_h)\overrightarrow{\tau}.$ However, with
respect to {\it space bracket:} $[\cdot{}_h,\cdot{}_h]_h$ torsion
of ${}^{(3)}\nabla$ is zero.} (relatively $h$) derivative on
$\mathcal{M}_h.$

Note also, that kinematic tensors $a,\omega$ and $\mathcal{D}$ are
generally covariant 4D tensors. This circumstance reveals
important  difference between reference frames and coordinate
systems: the fact that these tensor are nonzero does'nt depend of
choice of coordinate system, while by definition it is drastically
dependent on the choice of reference frame.

\subsection{Monads formalism in special gauge}

In spite of independency of kinematical tensors on the choice of
coordinate system, its apparent representation can be more or less
complicated in the coordinate systems, differently adopted to the
$\tau$-field.  Concrete  realizations  of monads formalism in
special coordinate systems  is called {\it gauges of monads
formalism}. In a difference from a pure mathematical formalism,
which development is more compact and convenient when one is based
on 1-form $\tau,$ using of the vector field
$\overrightarrow{\tau}$ is more preferable in the most of applied
problems. Physical (or geometrical) cause of this preference is
due to the fact, that 1-form $\tau$ defines local simultaneity
space, which we are commonly don't deal with in our experiments.
From the other hand law of motion of a  particles, forming some
reference frame and vector field $\overrightarrow{\tau}$ (this is
4-velocity field of the particles) are very often available for
experimental observations. Mathematically the difference between
fields $\tau$ and  $\overrightarrow{\tau}$ is caused by the
theorem stating possibility of straightening of the vector fields
and by absence of the similar theorem  stating possibility of
"straightening"\, of the 1-forms.

Let consider in more details monads formalism in the coordinate
system, which straights  $\tau$-field, i. e. in which we have
\begin{equation}\label{chr}
\overrightarrow{\tau}=\frac{\partial}{\partial t}.
\end{equation}
Such gauge is particular case  of the so called chronometric gauge
of monads method \cite{1}, for which
\begin{equation}\label{chr1}
\overrightarrow{\tau}=\frac{1}{\sqrt{g_{00}}}\frac{\partial}{\partial
t}.
\end{equation}
We'll call this very special gauge (\ref{chr1})  {\it strong
chronometric gauge} or more briefly {\it canonical gauge}. In a
difference with classical chronometric gauge canonical one
assumes, that not only lines of time coincide with coordinate
lines $x^i=\text{const},$ $(i=1,2,3),$ but also the coordinate
$x^0=t$ is world (i.e. global) time. Note, that the choice of
canonical gauges bring no any restrictions on the type of
reference frame or on the metric of the manifolds. This choice
only assumes fixation of  some definite coordinate system on the
manifold.

In the canonical gauge the metric $g$ can be written in the form:
\begin{equation}\label{metrchr}
g=dt\otimes dt+\Omega_i dt\vee dx^i-\sum\limits_{i=1}^3H_i^2dx^i\otimes dx^i-\sum\limits_{i=1}^3\delta_i(dx\vee
dx)^i,
\end{equation}
where
\[
\Omega_i\equiv g_{0i},\quad H^2_i\equiv -g_{ii},\quad \delta_1\equiv-g_{23};\ \delta_2\equiv-g_{13};\
\delta_3\equiv-g_{12};\quad (dx\vee dx)^1\equiv dx^2\vee dx^3,\dots
\]
etc.  From the  (\ref{metrchr}) we obtain
\begin{equation}\label{tauchr}
\tau=dt+\Omega_idx^i,
\end{equation}
and for space metric by (\ref{dec})
\begin{equation}\label{hmetrchr}
h=\sum\limits_{i=1}^3(H_i^2+\Omega_i^2)dx^i\otimes dx^i+\sum\limits_{i=1}^3(\delta_i+(\Omega\Omega)_i)(dx\vee
dx)^i,
\end{equation}
where $(\Omega\Omega)_1=\Omega_2\Omega_3,$
$(\Omega\Omega)_2=\Omega_1\Omega_3,$
$(\Omega\Omega)_3=\Omega_1\Omega_2.$

Finally by (\ref{chr}) and space-like nature of kinematic tensors
their components are satisfied to the equations:
\begin{equation}\label{cond1}
a_0=0;\quad \omega_{0i}=\omega_{i0}=0;\quad
\mathcal{D}_{i0}=\mathcal{D}_{0i}=0.
\end{equation}

\subsection{Kinematical tensors and observables}

Let us take our attention to the physical sense of kinematical
tensors and their relations to observable values. Since results of
our discussion will be used in searching for metrics, it is
important from the beginning to clear true geometrical nature of
the physical values, that are described by the kinematical tensors
\cite{koka}. Geometrically  4-acceleration is 4-vector. Writing
this vector in the form:
\begin{equation}\label{acc1}
\overrightarrow{a}=a^\alpha\partial_\alpha
\end{equation}
and using special properties of canonical gauge, we go to the
following kind of space-likeness condition for  the vector
$\overrightarrow{a}$:
\begin{equation}\label{acc2}
g(\overrightarrow{a},\overrightarrow{\tau})=0\Rightarrow
a^0+\Omega_ia^i=0,
\end{equation}
that leads to the expression
\begin{equation}\label{acc3}
\overrightarrow{a}=-(\Omega_ia^i)\partial_t+a^i\partial_i.
\end{equation}
Observable quantities are the so called physical components $\bar
a^i,$ which are connected with 3D coordinate components  $a^i$ by
the well known relations\footnote{In fact we use affine
(non-orthogonal and non-normalized) coordinate tetrad fields
$\{\partial_t,\partial_1,\partial_2,\partial_3\}$. By the fact
physical components contain multipliers
$\sqrt{g_{ii}}=\|\partial_i\|,$ rather than
$\sqrt{h_{ii}}=\|(\partial_i)_h\|.$ The difference of descriptions
of space values in terms  $\{\partial_i\}$ and
$\{(\partial_i)_h\}$ has no any influence on final conclusions.
Moreover, in what follows we'll not  need  complete tetrad
reference frame formalism. }:
\begin{equation}\label{acc4}
\bar a^i=\sqrt{g_{ii}}a^i=H_ia^i\quad\text{(no summation!).}
\end{equation}
%(используем пространственную метрику $h$ для пространственных векторов и
%тензоров).
So, we obtain the following final expression for the
4-acceleration vector in terms of observable:
\begin{equation}\label{acc5}
\overrightarrow{a}=-\left(\sum\limits_{i=1}^3\frac{\Omega_i\bar a^i}{H_i}\right)\partial_t+\sum\limits_{i=1}^3\frac{\bar a^i}{H_i}\partial_i.
\end{equation}
The acceleration 4-covector has the following kind in terms of
observable quantities  $\bar a^i$ (manipulations with indexes can
be yield by the metric (\ref{hmetrchr}), taken with opposite
sign):
%\begin{equation}\label{acc5a}
\[
a=-\left(\frac{H_1^2+\Omega_1^2}{H_1}\bar a^1+\frac{\delta_3+\Omega_1\Omega_2}{H_2}\bar a^2+\frac{\delta_2+\Omega_1\Omega_3}{H_3}\bar
a^3\right)\,dx^1+
\]
\begin{equation}\label{acc6}
-\left(\frac{\delta^3+\Omega_1\Omega_2}{H_1}\bar a^1+\frac{H_2^2+\Omega_2^2}{H_2}\bar a^2+\frac{\delta_1+\Omega_2\Omega_3}{H_3}\bar
a^3\right)\,dx^2+
\end{equation}
%\begin{equation}\label{acc7}
\[
-\left(\frac{\delta_2+\Omega_1\Omega_3}{H_1}\bar a^1+\frac{\delta_1+\Omega_2\Omega_3}{H_2}\bar
a^2+\frac{H_3^2+\Omega_3^2}{H_3}\bar a^3\right)\,dx^3.
%\end{equation}
\]

The angular velocity of continuum media is in its geometrical
nature bivector\footnote{This statement is supported by the
consideration that angular velocity will remain bivector in space
of any dimensions.} of 3D space, i.e. this is  3D antisymmetric
contravariant tensor of the kind
\begin{equation}\label{avel1}
\dot\varphi\equiv\varphi^{ik}\partial_i\wedge\partial_k.
\end{equation}
In practice the well known isomorphism between vectors and
bivectors in 3D space allows to consider angular velocity as the
vector  $\overrightarrow{\omega},$ which direction defines instant
axe of rotation and absolute value defines the number of radians,
that rotating elements goes per unit of time. Transition from the
vector $\overrightarrow{\omega}$ to the space spin tensor
$\omega,$ defined as above, is given by the relation:
\begin{equation}\label{avel2}
\omega=\pi_{\overrightarrow{\tau}}(\ast\overrightarrow{\omega})=\star\overrightarrow{\omega},
\end{equation}
or in the components
\begin{equation}\label{avel3}
\omega_{\alpha\beta}=\sqrt{|\Delta|}\varepsilon_{\gamma\delta\alpha\beta}\tau^\gamma\omega^{\delta}
=\sqrt{|\Delta|}\varepsilon_{0i\alpha\beta}\omega^{i},
\end{equation}
where the special kind of $\overrightarrow{\tau}$ in canonic gauge
have been used. Going to the physical components $\bar \omega^i,$
we finally obtain:
\begin{equation}\label{avel4}
\omega_{ik}=\sqrt{|\Delta|}\sum\limits_{s=1}^3\varepsilon_{ik0s}\frac{\bar\omega^{s}}{H_s}.
\end{equation}
Interpretation of rate of strain tensor is based on its
contribution to deformation of space metric along the flow of
world lines of reference frame (the third formula in
(\ref{formull})). The components of covariant rate of strains
tensor are directly connected with the rates of relative
deformations of the lengthes, measured along coordinate lines, and
with the rate of variations of the angles between the coordinate
lines \cite{koka2}. In the canonical gauge the following relations
between physical components of rate of strains tensor and its
coordinate components takes place
\begin{equation}\label{def1}
\mathcal{D}_{ik}=\frac{\bar{\mathcal{D}}_{ik}}{\sqrt{g^{ii}}\sqrt{g^{kk}}}.
\end{equation}

\section{Kinematical statement of the problem:\\ reconstruction of a metric from the $\tau$-field\\ and kinematical tensors}

Let consider some reference frame, defined by the
$\overrightarrow{\tau}$-field and let prescribe to the physical
components of kinematic tensors, related to this reference frame,
some arbitrary values. Without loss of generality reference frame
and kinematical tensors can be considered in canonical gauge.
Moreover such gauge is the most natural for the observers,
attached to material elements of the reference frame. So, we have
 $\overrightarrow{\tau}$-field,
given by the formula  (\ref{chr}) and the set of 12 values $\{\bar
a^1,\bar a^2,\bar a^3,\bar \omega^1,\bar \omega^1,\bar
\omega^1,\bar D_{11},\bar D_{12},\bar D_{13},\bar D_{22},\bar
D_{23},\bar D_{33},\}$ defined by the formulas  (\ref{acc6}),
(\ref{avel2}) and (\ref{def1}), every of which contains as unknown
values the components of metric  $g$ to be defined in future.
Expression (\ref{decv}), defining kinematic tensors, one should to
consider now as equations for metric. In view of (\ref{chr}) left
hand side of expression (\ref{decv}) takes the form:
\begin{equation}\label{decv1}
\nabla_\alpha\tau^\beta=\Gamma^\beta_{\alpha0},
\end{equation}
so the expression  (\ref{decv}) can be rewritten in the following
equivalent form:
\begin{equation}\label{decv2}
\frac{\partial g_{\gamma0}}{\partial x^\alpha}+\frac{g_{\alpha\gamma}}{\partial
t}-\frac{\partial g_{\alpha0}}{\partial
x^\gamma}=2(\tau_{\alpha}a_{\gamma}+\mathcal{H}_{\alpha\gamma}).
\end{equation}
The components of equation (\ref{decv2}) with $\alpha=0,$
$\gamma=0$ and  $\alpha=i,$ $\gamma=0$ are satisfied identically
by gauge conditions. Nontrivial equations are obtained for the
components  with $\alpha=0,$ $\gamma=i$
\begin{equation}\label{eq1}
\frac{\partial\Omega_i}{\partial t}=a_i
\end{equation}
and with $\alpha=i,$ $\gamma=k$
\begin{equation}\label{eq2}
\frac{\partial g_{ik}}{\partial t}+\frac{\partial\Omega_k}{\partial
x^i}-\frac{\partial\Omega_i}{\partial
x^k}=2(\Omega_i a_k+\mathcal{H}_{ik}).
\end{equation}
Decomposing symmetric and antisymmetric part in the latter
equation, we go to the following pair equations:
\begin{equation}\label{eq3}
\frac{\partial g_{ik}}{\partial
t}=2\mathcal{D}_{ik}+\Omega_ia_k+\Omega_ka_i,
\end{equation}
and
\begin{equation}\label{eq4}
\frac{\partial\Omega_k}{\partial
x^i}-\frac{\partial\Omega_i}{\partial
x^k}=2\omega_{ik}+\Omega_ia_k-\Omega_ka_i,
\end{equation}
that is concordant with coordinateless definitions (\ref{defexc})
and (\ref{deftors}). The system of equations  (\ref{eq1}),
(\ref{eq3}) and (\ref{eq4}) will be a subject of further
investigations.

\subsection{Integrability condition}

The remarkable circumstance of the subsystem (\ref{eq1}),
(\ref{eq3}), defining time dependency of metric components, is its
conditionless  integrability. Really, space coordinates, which
metric components and kinematic tensors depend on, can be viewed
in the system as parameters. So, the system
(\ref{eq1}),(\ref{eq3}) can be treated as ordinary differential
equations system of the first order, resolved with respect to the
first derivatives on $t$. In view of general theorems of the
theory of differential equations solution to the system
(\ref{eq1}), (\ref{eq3}) locally exists and depends on 9
"integration constants"\,, which in fact are "functional
constants"\, i.e. some undefined functions of space coordinates.

Some problems with integrability can arise only for the equation
(\ref{eq4}), which partially define dependency of the metric on
space variables. For the analysis of integrability of the equation
let's introduce the following space 1-forms:
\[
\Omega\equiv \Omega_i\,dx^i;\quad a\equiv a_i\,dx^i;\quad \omega=\omega_{ik}\,dx^i\wedge dx^k
\]
and let's define sectional space external differential:
\[
\not\! d\equiv d|_{t=\text{const}}.
\]
Then the subsystem  (\ref{eq1}),(\ref{eq4}) can be rewritten more
compactly:
\begin{equation}\label{int1}
\dot\Omega=a;\quad \text{(I)}\quad\quad \not\! d\,\Omega=2\omega+\Omega\wedge
a.\quad
\text{(II)}
\end{equation}
Hereafter the dot over letter will denote standard partial
derivative on  $t.$ The equation (\ref{int1}.II) means, that
righthand side is exact space form. In view of nilpotency of
external differentiation ($\not\! d\ \circ\not\! d\equiv 0$) the
necessary integrability condition for  (\ref{int1}.II) (locally
this is also sufficient condition!) is closeness condition for the
righthand part with respect to  $\not\! d$:
\begin{equation}\label{int2}
\not\! d(2\omega+\Omega\wedge
a)=0,
\end{equation}
which must be satisfied as a consequences of original equations.
Using the well known rules of wedge product and differentiation,
commutativity of  $\partial_t$ and $\not\! d$ and equations
(\ref{int1}), we go to the following chain of equalities for
(\ref{int2}):
\[
2\not\! d\omega+\not\! d(\Omega\wedge\dot\Omega)=2\not\! d\omega+ \not\!
d\Omega\wedge\dot\Omega-\Omega\wedge\not\! d\dot\Omega=2\not\!
d\omega+2\omega\wedge\dot\Omega-\Omega\wedge(\not\! d\Omega)^{\cdot}=
\]
\begin{equation}\label{int3}
2(\not\!
d\omega+\omega\wedge\dot\Omega-\Omega\wedge\dot\omega)=0.
\end{equation}
In fact the integrability condition (\ref{int3}) allows generally
covariant formulation, if one will start from the equation
(\ref{deftors}). Rewritten it in the form
\begin{equation}\label{int4}
d\tau=2\omega+\tau\wedge a
\end{equation}
and applying differential to the both sides, we go to the
closeness condition of 2-forms in the righthand side of
(\ref{int4}) in the following form:
\begin{equation}\label{int5}
d(2\omega+\tau\wedge a)=0.
\end{equation}
Taking into account the identity
\begin{equation}\label{int6}
\tau=dt+\Omega,\quad d\lambda=(-1)^s\dot\lambda\wedge dt+\not\!
d\lambda,
\end{equation}
where $s$ --- is rank of external form, it is easily to show, that
(\ref{int5}) is equivalent to (\ref{int3}).

Lets make some remarks to the obtained results.
\begin{enumerate}
\item
Integrability condition (\ref{int5}) in the form of vanishing of
some 3-form in 4D space-time after applying of dualization
operator would be equivalent to vanishing of some 4-vector i.e.
four conditions on kinematical tensors. Canonical gauge reveals,
that we are dealt with space 3-form  in (\ref{int5}) and 3D
dualization $\star$ leads only to one essential condition.
\item
Rather simple calculations with $\star$-operator, applied to
(\ref{int3}), lead to the following form of integrability
condition:
\begin{equation}\label{int7}
{}^{(3)}\text{div}\,\overrightarrow{\omega}+2a(\overrightarrow{\omega})-2\Omega(D_t\overrightarrow{\omega})=0,
\end{equation}
where
\[
{}^{(3)}\text{div}\equiv\star\not\!
d\,\star;\quad D_t\equiv\star\partial_t\star
\]
or in components:
\begin{equation}\label{int8}
\frac{1}{\sqrt{|\Delta|}}\partial_i(\sqrt{|\Delta|}\omega^i)+2a_i\omega^i-2\frac{\Omega_i}{\sqrt{|\Delta|}}\frac{\partial}{\partial
t}(\sqrt{|\Delta|}\omega^i)=0.
\end{equation}
\item
For the formula (\ref{deftors}), treated within the context of
standard geometrical statement of the problem on kinematical
tensors calculation, the condition (\ref{int5}) or equivalent
condition (\ref{int3}) are identities and they are satisfied due
to definitions of the kinematical tensors. In our kinematical
statement of the problem these relations must be interpreted in
other manner. In order to avoid additional equations on metric
components and chain of integrability conditions, {\it we should
consider relation (\ref{int7}) or its equivalent forms as
restrictions on the components of kinematical tensors $\omega$ and
$a$, or, more exactly, as a restrictions on its physical
components.} In other words, in general, kinematical tensors
$\omega$ and  $a$ can't be given absolutely independently from
each other and from geometry, but they must satisfy (\ref{int7})
in special coordinate system. Connection between tensors $\omega$,
$a$ and the metrics, expressed by (\ref{int7}), should be taken
into account by some way, for example, by resolving of
(\ref{int7}) with respect to one of the physical components of the
tensors.
\item
Note, that integrability condition (\ref{int7}) become identity
under  $\omega=0.$ From the other hand, this equation is necessary
and sufficient condition of integrability of 1-form of time, as it
directly follows from (\ref{int4}) and the well known Frobenius
theorem. Integrable distribution $\ker\tau$ defines local
foliation of space-time on submanifolds of simultaneous events,
parametrized by $t.$ So, the fact  of nontrivial relation of
kinematics and geometry is in closest relation with non-holonomic
aspects of  submanifolds of simultaneous events in case of
$\omega\neq0.$
\item
Physically the relation (\ref{int7}) generalizes nonrelativistic
3D hydrodynamical formula $\text{div}\overrightarrow{\omega}=0$
for the case of relativistic motions on 3D Riemannian
non-holonomic submanifolds.
\item
Mathematically relation  (\ref{int7}) is consequence of
commutativity of space partial derivatives. The condition of
commutativity of the space partial derivatives with time one can
be easily obtained from  (\ref{int2}) by differentiating with
respect to $t$ and it does'nt produce any new independent
conditions.
\item
The components of rate of strain tensor does'nt appear in apparent
form in (\ref{int7}), so it can be prescribed arbitrarily.
\item
In case of satisfying of (\ref{int7}) the reconstructed metric $g$
will have considerable functional arbitrariness. It becomes clear
after calculation of number of functions and constraints. In case
of nonzero spin it will be  $12-1=11$ (total number of components
rotation, acceleration and deformation minus one restriction).
From the other hand, there are only 9 unknown components of
metric, which are to be found (the component $g_{00}=1$ due to
choosen gauge), from which only 6 are physically essential in view
of possibility of four general coordinate transformation (that may
violate canonical gauge). In absence of rotation we have 12
arbitrary functions against 6 physically essential components of a
metric.
\end{enumerate}

\section{Example 1: space-times with stationar field\\ of acceleration}

Let consider the reference frame in canonic gauge, for which
$\mathcal{D}=0,$ $\omega=0$ and $a\neq0.$ One of the space
coordinate line (for example $x^1$) can always be
adopted\footnote{Let us remind, that after canonical gauge fixing
we have allowable coordinate transformation of the form:
\[
{x'}^0=x^0+\psi(x^1,x^2,x^3);\quad {x'}^i=\chi^{i}(x^1,x^2,x^3),
\]
where $\psi$ and $\chi^i$ $(i=1,2,3)$ --- arbitrary smooth
functions, providing invertibility of the transformations. Using
this remained coordinate degrees of freedom, one can transform
acceleration field to the form mentioned in the text.} to the
acceleration field so, that $\overrightarrow{a}=a^1\partial_1.$
We'll denote the  nonzero physical component of acceleration as
$\bar a^1 \equiv\bar a.$ Then with using (\ref{acc6}) we go to the
following expression for 1-form of acceleration:
\begin{equation}\label{ac1}
a=-\frac{H_1^2+\Omega_1^2}{H_1}\bar a\, dx^1-\frac{\delta_3+\Omega_1\Omega_2}{H_1}\bar a\,
dx^2-\frac{\delta_2+\Omega_1\Omega_3}{H_1}\bar a\,
dx^3.
\end{equation}
Let us go to the new unknown functions:
\[
X_i\equiv\frac{\Omega_i}{H_i};\quad
\Delta_i\equiv\frac{\delta_i}{(\Omega\Omega)_i}\quad (i=1,2,3).
\]
In the variables  $X_i,\Delta_i$ the subsystem of equations
(\ref{eq1}), (\ref{eq3}), defining time dependency of the metric,
takes the following form:
\begin{equation}\label{s1}
\dot X_1=-(1+X_1^2)^2\bar a;\quad \dot
X_2=-(1+X_2^2)(\Delta_3+X_1X_2)\bar a;\quad \dot
X_3=-(1+X_3^2)(\Delta_2+X_1X_3)\bar a;
\end{equation}
\begin{equation}\label{s2}
\dot \Delta_1=[X_2(\Delta_2-\Delta_1\Delta_3)+X_3(\Delta_3-\Delta_1\Delta_2)-X_1(\Delta_1X_2^2+\Delta_1X_3^2-2X_2X_3)]\bar a;
\end{equation}
\begin{equation}\label{s3}
\dot \Delta_2=[(\Delta_2+X_1X_3)(X_1-\Delta_2X_3)+(1+X_1^2)(X_3-\Delta_2X_1)]\bar a;
\end{equation}
\begin{equation}\label{s4}
\dot \Delta_3=[(\Delta_3+X_1X_2)(X_1-\Delta_3X_2)+(1+X_1^2)(X_2-\Delta_3X_1)]\bar a.
\end{equation}
By using new variables we have reduced the system of a nine
ordinary differential equations (\ref{eq1}), (\ref{eq3}) to the
system of a six equations  (\ref{s1})-(\ref{s4}). The excluded
functions $H_i$ are expressed through the variables
$X_i,\Delta_j$ as follows:
\begin{equation}\label{s5}
(\ln H_1)^\cdot=X_1(1+X_1^2)\bar a;\quad (\ln H_2)^\cdot=X_2(\Delta_3+X_1X_2)\bar
a;\quad (\ln H_3)^\cdot=X_3(\Delta_2+X_1X_3)\bar a;.
\end{equation}

Let consider particular case of the system (\ref{s1})-(\ref{s4}),
admitting complete integration: $\Delta_1=\Delta_2=\Delta_3=0,$
$\dot{\bar a}=0.$ Substituton of the first three conditions into
the system  (\ref{s1})-(\ref{s4}) leads to the integrals:
$X_2=X_3=0$ and to the following unique equation:
\begin{equation}\label{s6}
\dot X_1=-(1+X_1^2)^2\bar a,
\end{equation}
which under the imposed stationarity condition can be easily
integrated by separation of variables:
\begin{equation}\label{s7}
-\bar at+F(x^1,x^2,x^3)=\frac{X_1}{2(1+X_1^2)}+\frac{1}{2}\arctan
X_1,
\end{equation}
where $F$ is yet unknown function of space coordinates.
Integrating (\ref{s5}), we go to the following expression for Lame
coefficients:
\begin{equation}\label{s8}
H_1=\frac{\Phi_1(x^1,x^2,x^3)}{\sqrt{1+X_1^2}};\quad
H_2=H_2(x^1,x^2,x^3);\quad H_3=H_3(x^1,x^2,x^3),
\end{equation}
where $\Phi_1$ is unknown function of space coordinates, $H_2,H_3$
are arbitrary functions of space coordinates. For the unique
nonzero component $\Omega_1$ we obtain the following expression:
\begin{equation}\label{s9}
\Omega_1=H_1X_1=\frac{\Phi_1(x^1,x^2,x^3)X_1}{\sqrt{1+X_1^2}}.
\end{equation}
The equation (\ref{eq4}) gives two additional restrictions for
dependency of  $\Omega_1$ on space coordinates:
\begin{equation}\label{s10}
\frac{\partial\Omega_1}{\partial x^2}=\frac{\partial\Omega_1}{\partial
x^3}=0.
\end{equation}
These equation will be identically satisfied by the following
dependencies:
\begin{equation}\label{s11}
\bar a=\bar a(x^1);\quad F=F(x_1);\quad \Phi_1=\Phi(x^1).
\end{equation}
Without loss of generality the function $\Phi(x^1)$ can be put to
1 by suitable choice of the coordinate  $x^1,$ which after
imposing of all restrictions is defined up to arbitrary
diffeomorphism ${x'}^1=f(x^1).$ Now the obtained metric can be
written in the form:
\begin{equation}\label{s12}
g=dt\otimes dt+\tanh\psi(dt\otimes dx^1+dx^1\otimes
dt)-(1-\tanh^2\psi)dx^1\otimes dx^1-
\end{equation}
\[
H_2^2(x^1,x^2,x^3)dx^2\otimes
dx^2-H_3^2(x^1,x^2,x^3)dx^3\otimes dx^3,
\]
where the function  $\psi$ is defined by the relation:
\begin{equation}\label{s13}
\frac{1}{2}\frac{\tanh\psi}{\cosh\psi}+\arctan e^{\psi}=\bar
a(x^1)t+F(x^1).
\end{equation}
So, the class of space-times (\ref{s12}) prescribes to the
reference frames of the kind $\tau=\partial_t$ zero kinematical
tensors $\omega$ and $\mathcal{D},$ while  $\bar a^1=\bar a(x^1),$
where $\bar a(x^1)$ is arbitrary function of $x^1,$ defining the
field of physical components of acceleration vector.

\section{Example 2: space-times with isotropic\\ field of deformations}

In present section we are going to derive general kind of
space-times, having canonical reference frames  with  $a=0,$
$\omega=0$ and
\begin{equation}\label{d1}
\mathcal{D}=\sigma h,
\end{equation}
where $\sigma$ is arbitrary scalar function of all four
coordinates of the manifold. The deformations field (\ref{d1})
describes isotropic but non-homogeneous and non-static deformation
of the reference frame at every point of the manifold. Assuming
$\Omega=0,$ we find, that in the system  (\ref{eq1})-(\ref{eq4})
the only equations  (\ref{eq2}) remain non-trivial. They take the
following form:
\begin{equation}\label{d2}
\dot H_i=-\sigma H_i;\quad \dot\delta_i=-2\sigma\delta_i.
\end{equation}
These equations can be easily integrated and the solution has the
following form:
\begin{equation}\label{d3}
g=dt\otimes dt-e^{-2\int\sigma\, dt}\left(\sum\limits_{i=1}^3H_{0i}^2dx^i\otimes dx^i-\sum\limits_{i=1}^3\delta_{0i}(dx\vee
dx)^i\right),
\end{equation}
where $H_{0i},\delta_{0i}$ are arbitrary functions of space
coordnates.  The metric (\ref{d3}) generalizes the well known
Friedman-Robertson-Walker cosmological metrics for non-homogeneous
case.

\section{Example 3: the space-times with stationary rotated reference frames}

Finally let us consider one simple class of space-times, admitting
the canonical reference frames with stationary rotation. Assuming
in the equations (\ref{eq1})-(\ref{eq4}) $a=0,$ $\mathcal{D}=0,$
$\omega^2=\omega^3=0,$ $\omega^1\neq0,$ $\dot\omega^1=0,$ we go to
the following stationarity condition for metric:
$g_{\alpha\beta}=g_{\alpha\beta}(x^1,x^2,x^3).$ Using this
condition, one can transform integrability condition (\ref{int8})
to the following simple kind:
\[
\partial_1\left(\sqrt{|\Delta|}\frac{\bar\omega^1}{H_1}\right)=0,
\]
wherefrom
\[
\bar\omega^1=\frac{H_1}{\sqrt{|\Delta|}}\psi(x^2,x^3),
\]
where $\psi$ is arbitrary function of the pair of coordinates
$x^2,x^3.$ Substituting the last expression into the equation
(\ref{eq4}) and assuming $\Omega_1=0,$ we get:
\begin{equation}\label{r1}
\Omega_2=\Omega_2(x^2,x^3);\quad \Omega_3=\Omega_3(x^2,x^3);\quad
\frac{\partial\Omega_3}{\partial
x^2}-\frac{\partial\Omega_2}{\partial x^3}=2\psi(x^2,x^3).
\end{equation}
Up to the unessential addends, that are proportional to some exact
form\footnote{The equation (\ref{eq4}) defines  $\Omega$ up to an
arbitrary exact  form, which can be compensated by purely space
coordinate transformations.} the solution takes the form:
\begin{equation}\label{r2}
\Omega_2=-\int\psi\, dx^3;\quad \Omega_3=\int\psi\, dx^2.
\end{equation}
Finally, the class metric, admitting stationary rotated rigid
geodesic canonical reference frame, is described by the
expression:
\begin{equation}\label{r3}
g=dt\otimes dt-\left(\int\psi(x^2,x^3)\, dx^3\right)(dt\otimes
dx^2+dx^2\otimes dt)+\left(\int\psi(x^2,x^3)\, dx^2\right)(dt\otimes
dx^3+dx^3\otimes dt)
\end{equation}
\[
+ g_{ik}(x^1,x^2,x^3)dx^i\otimes dx^k,
\]
where $\psi$ and $g_{ik}$ are arbitrary function of the written
coordinates.

\section{The mixed (kinematically-geometrical) problem}

Now let us consider the situation, when the reference frame is
unknown, while metric and kinematical tensors are forgiven. In
such problem the concordance conditions for the metric and
kinematical tensors are much more rigid, then in previous
problems. Really, the fact, that kinematical tensors are different
projection of covariant derivative of monads field\footnote{Since
the metric now is forgiven, one may do not take care on differing
of co- and contravariant versions of kinematical tensors. Now it
will be more conveniently for us to be dealt with the field of
1-form $\tau$ and covariant representation of kinematical
tensors.} leads to the very complicated  system of integrability
conditions, following from the basic relation:
\begin{equation}\label{mix1}
[\nabla,\nabla]\tau=-\hat R(\tau),
\end{equation}
where $\hat R$ is curvature operator. Moreover, purely algebraic
restrictions arise for the kinematical tensors themselves: in view
of the fact, that the field  $\tau$ must be orthogonal to all
three kinematic tensors:
\begin{equation}\label{mix2}
a(\overrightarrow{\tau})=0;\quad \omega(\ ,\overrightarrow{\tau})=0;\quad
\mathcal{D}(\ ,\overrightarrow{\tau})=0,
\end{equation}
and the fact of non-degeneracy of a metric, the following
condition must be satisfied identically:
\begin{equation}\label{mix3}
\text{rank}(a,\omega,\mathcal{D})^T\le3,
\end{equation}
where the braces denote (composed) matrix of the system
(\ref{mix2}), which one can consider as the system of linear
equations with respect to components $\tau^\alpha.$ The condition
(\ref{mix3}) expresses the necessary condition of non-trivial
compatibility for this system. The second necessary condition is
requirement that the space of solutions
$\ker(a,\omega,\mathcal{D})^T$ must include time-like direction.
Even these condition are in fact  very strong: some "randomly
written"\, system of kinematical tensors under fixed metric will
not satisfy these conditions.

In order to analyze integrability conditions, following from the
(\ref{mix1}), we calculate commutator from the left in
(\ref{mix1}) using the definition\footnote{It will be more
convenient for us to collect the sum $\omega+\mathcal{D}$ into one
space-projected tensor $\mathcal{H}.$} (\ref{decf}) and substitute
the result into the (\ref{mix1}). After some algebra the result
takes the following form:
\begin{equation}\label{mix4}
-[\hat R +(\widehat{\text{Id}}\wedge a)\otimes a-(\widehat{\text{Id}}\wedge
\nabla)a](\tau)=2\omega\otimes a+\nabla\wedge \mathcal{H}
\end{equation}
or in the components
\begin{equation}\label{mix5}
-[\hat R^{\gamma}_{\cdot\sigma\alpha\beta} +(\delta^\gamma_\alpha a_\beta-\delta^\gamma_\beta
a_\alpha)a_\sigma+(\delta^\gamma_\beta\nabla_\alpha-\delta^\gamma_\alpha\nabla_\beta)a_\sigma]\tau_\gamma=2\omega_{\alpha\beta}a_\sigma+
\nabla_\alpha
\mathcal{H}_{\beta\sigma}-\nabla_\beta\mathcal{H}_{\alpha\sigma}.
\end{equation}
The equations (\ref{mix5}) can be understood as strongly
overdefined system of linear equations  (in general their number
will be 24) with respect to the components $\tau_\alpha.$ Its
non-trivial compatibility implies coincidence of the ranks of
matrix and  enlarged matrix (Kronecker-Capelli theorem).
Differently from the algebraic conditions (\ref{mix3}), the
conditions on the ranks for  (\ref{mix5}) will involve derivatives
of kinematical tensors. So, kinematical tensors and metric are
concordant to each other much more rigidly, then kinematical
tensors and  $\tau$-field, in spite of the first are commonly
defined through the  $\tau$-field by relation (\ref{decf}).

\section{Undeformed geodesic reference frames in spherically-symmetric space-times}\label{spher}

As example let us consider the problem of finding of canonical
reference frames with $\mathcal{D}=0,$ $a=0,$ $\omega\neq0$ in
space-times of the following kind:
\begin{equation}\label{sph1}
g=e^{2\nu(t,r)}dt\otimes dt-e^{2\lambda(t,r)}dr\otimes dr-r^2(d\theta\otimes d\theta+\sin^2\theta d\varphi\otimes
d\varphi).
\end{equation}
The metric (\ref{sph1}) describes geometry of spherically
symmetric space-time in special coordinates (co-moving curvature
coordinates). The condition  $a=0$ means that vector field
$\overrightarrow{\tau}$ is geodesic and satisfies the equation
$\nabla_{\overrightarrow{\tau}}\overrightarrow{\tau}=0.$ The
condition $\mathcal{D}=0$ with the definition (\ref{defexc}) and
previous condition mean, that the field $\overrightarrow{\tau}$ is
Killing's field, i. e. satisfies Killing's equations:
\begin{equation}\label{sph2}
L_{\overrightarrow{\tau}}g=0
\end{equation}
and is element of isometry algebra of space-time with the metric
(\ref{sph1}).

We'll describe the field  $\overrightarrow{\tau}$ by its
coordinate components:
\[
\overrightarrow{\tau}=\tau^t\partial_t+\tau^r\partial_r+\tau^\theta\partial_\theta+\tau^\varphi
\partial_\varphi,
\]
which are restricted by the following normalizing condition:
\begin{equation}\label{normmm}
e^{2\nu}(\tau^t)^2-e^{2\lambda}(\tau^r)^2-r^2(\tau^\theta)^2-r^2\sin^2\theta\,(\tau^\varphi)^2=1.
\end{equation}

By technical considerations it will be more conveniently for us to
derive integrability conditions directly from the equations for
components of kinematical tensors, rather then from the equations
(\ref{mix4}). In view of linearity of Killing's equations
(\ref{sph2}) it is conveniently to begin analysis namely from
these equations. In the components the  Killing's equations take
the following kind:
\[
\tau^t\dot\nu+\tau^r\nu'+\dot\tau^t=0;\ (1)\ \quad
\tau^t\dot\lambda+\tau^r\lambda'+{\tau^r}'=0;\ (2)\
\]
\[
\tau^r+r\tau^\theta_{,\theta}=0;\ (3)\
\quad \frac{\tau^r}{r}+\cot\theta\,\tau^\theta+\tau^{\varphi}_{,\varphi}=0;\
(4)
\]
\[
e^{2\lambda}\dot\tau^r-e^{2\nu}{\tau^t}'=0;\ (5)\quad
r^2\dot\tau^\theta-e^{2\nu}\tau^t_{,\theta}=0;\ (6)\quad
r^2\sin^2\theta\,\dot\tau^{\varphi}-e^{2\nu}\tau^t_{,\varphi}=0; (7)
\]
\[
r^2{\tau^{\theta}}'+e^{2\lambda}\tau^r_{,\theta}=0;\ (8)\quad
r^2\sin^2\theta\,{\tau^{\varphi}}'+e^{2\lambda}\tau^r_{,\varphi}=0;\quad
(9)\
\sin^2\theta\,\tau^{\varphi}_{,\theta}+\tau^{\theta}_{,\varphi}=0.\ (10)
\]

Exact consecutive analysis of the system we bring out in Appendix.
Here we use finite result, expressed by the formulas
(\ref{taur})-(\ref{tauph}), (\ref{taut}) for the components
$\tau^\alpha.$ In view of the fact that
$\overrightarrow{\tau}$-field must be the field of  4-velocity of
reference frame, it should satisfy the normalizing condition
(\ref{normmm}). Substituting the formulas
(\ref{taur})-(\ref{tauph}), (\ref{taut}) into (\ref{normmm}) and
equating the coefficients under a different combinations of
trigonometric functions on the angles  $\theta$ and $\varphi$ we
go to the following simple deduction: {\it the unique time-like
unit Killing's field of space-time with spherical symmetry is the
field of the kind $\tau=\partial_t$ under  $\nu=0,$
$\dot\lambda=0$, i.e. for the case of special static metric.} This
space-time is direct sum  $R\oplus\mathcal{H}_3,$ where
$\mathcal{H}_3$ is arbitrary 3-dimensional manifold, possessing
SO(3)-symmetry with respect to some point.  It is easily to check
that the field $\partial_t$ in this space is geodesic and
non-rotating (it describes matter points, that are in a rest with
respect to 3-dimensional unmovable spherical coordinate system).
So,  {\it the unique class of spherically symmetric space-times,
admitting rigid geodesic reference frames are those with the
metric:}
\begin{equation}\label{soll}
g=dt\otimes dt-e^{2\lambda(r)}dr\otimes dr-r^2(d\theta\otimes d\theta+\sin^2\theta d\varphi\otimes
d\varphi).
\end{equation}
{\it The said reference frame is automatically non-rotating.}

\section{Conclusion}

\subsection{The geometrization principle}

Our analysis shows that the three components (metric, reference
frame and its kinematic characteristics) lying at the core of all
theoretical speculations of the observable phenomena within the
scope of the GR are interdependent to a degree. The standard
geometric statement of the problems means that the kinematic
characteristics of the reference frame can be calculated when the
$\tau$-field and metric are known. However, the practical
definition of either of these quantities depends, as a rule, on
considerable explicit or implicit a priori assumptions concerning
both theory behind the experiments being carried out and their
interpretation. Any attempt to determine the law of motion of the
reference frame or a space-time geometry will be based on the
accepted laws of electromagnetic signal propagation, on more or
less plausible hypotheses of the laws of distant body motion
(stars, galaxies and clusters), global and local properties of
space-time etc. Further analysis  of these problems discloses even
more a priori assumptions, for instance those which allow the
laboratory physical laws concerning extremal states and
characteristics of physical systems to be extrapolated. The
general picture is rather confounded, with observable values only
indirectly related to the estimable value. The most popular method
used to overcome these difficulties (probably fundamental flaws)
that emerged in the course of scientific practice is an attempt to
construct a consistent and comprehensive model of the Universe.
One of the attractions of the Einstein's gravity theory (and its
modern modifications) is that we can consistently describe, using
the same Einstein equations, both interior of the stars and the
global evolution of the Universe, the space-time scale of which
greatly exceeds the astrophysical one. On the other hand, the law
of free-fall
\begin{equation}\label{geom}
a=\nabla_{u}u=0
\end{equation}
(maybe with corrections due to the extended structure of test
bodies) describes the motion of the particles of cosmic dust,
planets, stars and galaxies. This law, slightly generalized (say,
in light of Kaluza-Klein theory) but basically the same in form
describes the motion of bodies and particles that have electrical
and other kinds of charge. On the basis of the geometrization
principle which has been introduced to physics as a result of a
successful development of the GR ideas, equation (\ref{geom}) is
declared to be fundamental and universal. It describes in its
ideal form the dynamics of any system by means of choosing
convenient metrics $g$ and (or) connection $\nabla$
 \cite{loshak,kok3}.

In this case geometry in the most general sense becomes a
"matrix"\, of sorts, matrix of the environment that exists in our
mind. However, in such situation  it loses "objectivity status".
Indeed, thanks to the wide array of tools employed by the modern
geometry, it appears almost invulnerable to the impact of
observable and experimental data. It becomes pretty apparent when
we look at the attempts to explain existing cosmological problems
(accelerated expansion, non-flat rotation curves of galaxies etc.)
in view of nonlinear $f(R)$-theories that are suggested today
\cite{odin,others}.

\subsection{Geometry or kinematics?}

This somewhat unexpected status of geometry shown by our study
brings to light another aspect that concerns the problem of
observables in GR. Any observation is being carried out within the
concrete reference frame and well-developed formalisms of
reference frames (monad's, tetrad's, spinor etc.) enable us to
calculate the observable effects with any degree of precision,
{\it provided the space-time geometry is known with the same
degree of precision.} But if the geometry is unknown or known only
approximately, or if we accept Poincar\'{e}'s point of view (see
discussion below) and consider geometry as a conditional entity to
be chosen more or less deliberately, methods of the theory of
reference systems lose their ability to predict results.

Let us illustrate the interaction between kinematics and geometry
and the nature of problems arising with the attempts to
distinguish one from the other in practice by example that refers
to theoretical interpretation of test particles motion. Let
observable law of the motion of free test particles be determined
by relation $x(t)=\{1,x^i(t)\},$ where $t$ is world time of the
canonical reference frame $\overrightarrow{\tau}=\partial_t.$ Let
this law of motion be geometrically interpreted within the scope
of a certain Riemannian geometry with metric $g$, such as the
dependency $x(t)$ is the solution of the geodesic equation in $t$-
parameterization:
\begin{equation}\label{z1}
\ddot x^\mu+\Gamma_{\lambda\sigma}^\mu\dot x^{\mu}\dot x^{\nu}=\frac{\ddot
s}{\dot s}\dot x^\mu,
\end{equation}
where
\begin{equation}\label{z2}
\dot s=\sqrt{g(\dot x,\dot x)};\quad \ddot s=\sqrt{(\dot x^\alpha\partial_\alpha g)(\dot x,\dot x)+2g(\ddot x,\dot
x)},
\end{equation}
and the point denotes  $t$-differentiation. It should be noted
that geometrization of the law of motion of test bodies, that is
fixation of $g$-metric at the same time determines all of the
kinematic characteristics of the canonical reference frame within
which the law has been observed. Let further more detailed
experiments with test bodies reveal a new law of motion of the
bodies in the same situation:
\begin{equation}\label{z3}
x'(t)=x(t)+\delta(t),
\end{equation}
with a minor correction $\delta(t).$ Purely geometric approach
consists of a suitable modification (correction) to geometry:
$g\to g'=g+\delta g,$ which  would account for observable
disturbances. Disturbance of $g$-metric will determine disturbance
$\delta$ of the  law of motion by means of second-order linear
differential equation obtained through linearization (\ref{z1}) in
the neighborhood of non-disturbed law of motion and metric:
\begin{equation}\label{z4}
\ddot\delta^\mu-\frac{\ddot s}{\dot s}\dot\delta^\mu+2\Gamma^{\mu}_{\lambda\sigma}\dot
x^\lambda\dot\delta^\sigma+(\delta\Gamma)_{\lambda\sigma}^\mu\dot x^{\mu}\dot
x^{\nu}-\frac{\ddot s}{\dot s}\dot\delta^\mu-\dot
x^{\mu}\frac{d}{dt}\left[\frac{(\delta^\alpha\partial_\alpha g+\delta g)(\dot x,\dot x)+2g(\dot x,\dot\delta)}{2\dot
s}\right]=0,
\end{equation}
where
\[
\delta\Gamma^\mu_{\lambda\sigma}=\frac{1}{2}g^{\mu\rho}(\delta g_{\rho\sigma,\lambda}+\delta g_{\rho\lambda,\sigma}-\delta
g_{\lambda\sigma,\rho})-
\delta g_{\alpha\beta} g^{\alpha\mu}\Gamma^{\beta}_{\lambda\sigma}.
\]

However, there is another (complementary) method to interpret
disturbance. It requires correction of the reference frame
characteristics while geometry remains constant. This view of the
law of motion disturbances means that the reference frame within
which the observations are being carried out is not quite
canonical and requires a minor additional transformation of
coordinates. At the same time it will compensate for the law of
motion disturbance and make the reference system canonical. In
order to determine the relation between the reference frame
modification $\delta\overrightarrow{\tau}$ and the law of motion
$\delta$ let us select $\delta\overrightarrow{\tau},$ such as the
transformation of the reference frame
$\overrightarrow{\tau}+\delta\overrightarrow{\tau}$ into a
canonical one is achieved by coordinate transformation related to
the disturbances of the law of motion $\delta.$ The requirement
that the $\tau$-field normalization is preserved leads in the
first order of smallness  to orthogonality of disturbances
$\delta\overrightarrow{\tau}$ to the basic
$\overrightarrow{\tau},$  therefore the disturbance of the
reference frame has only the space components $\delta\tau^i.$ If
we represent the desired coordinate transformations as:
\begin{equation}\label{coordtr}
t'=t,\quad {x'}^i=x^i+\xi^i(x),
\end{equation}
then, using the transformation law for vector components in the
first order of smallness, we obtain:
\[
{\tau'}^0=1;\quad {\tau'}^i=\frac{\partial\xi^i}{\partial
t}+\delta\tau^i,
\]
consequently, the requirement of canonicity $({\tau'}^i=0)$ leads
to the relation between the disturbance of the reference frame and
transformations that compensate for it:
\begin{equation}\label{tr1}
\xi^i=-\int\delta\tau^i\,dt +f^i(x^1,x^2,x^3).
\end{equation}

Let us now apply the obtained transformations to the disturbed law
of motion of probe particles. In the first order of smallness,
after having omitted purely space transformations related to
functions $f^i$ we obtain
\[
(x^i(t)+\delta^i(t))'=x^i(t)+\delta(t)+\xi=x^i(t)+\delta^i(t)-\int\delta\tau^i\,
dt.
\]
If we then select $\delta\tau^i$ such as the equations
\begin{equation}\label{comp}
\dot\delta^i=\delta\tau^i
\end{equation}
are true, we can achieve the desired result. Namely, that {\it the
corrections to the law of motion that come across in the course of
observations can be eliminated by means of transformation of the
reference frame within which these observations are being carried
out while the background geometry remains the same.}

Both approaches have been used in the course of the science
history. The GR and any geometrical gravity theory in general is
an example of a purely geometric solution to the problem of the
motion of test bodies. On the other hand, the Ptolemaic theory of
epicycles and deferents is essentially an example of a purely
kinematic solution to the problem of motion, based on the
Euclidian geometry of space and time. In light of our present
notions concerning the Universe we are forced to take an
intermediate point of view on the problem of observable motion,
especially as the details are being corrected all the time. In
these circumstances both the  geometry of ambient space-time and
the kinematics of the reference frame related to the Earth are
equally important for understanding of the observable motions.

\subsection{Poincar\'{e}'s conventionalism and its limitations}

Our analysis makes it clear that in the real situation correlation
(\ref{decf}) read as the definition of kinematic values is not
sufficient to describe the relation between the reference frame
kinematics and geometry. The fundamental equality of the geometric
and kinematic approaches discussed above comes most prominently
into view if we turn to the general approach to geometry,
suggested by A. Poincar\'{e}.

"Where do the initial principles of geometry come from? Are they
dictated by logic? By founding the non-Euclidean geometries
Lobachevsky established that is not the case. Do we discover
spaces with the help of the senses? The answer is no once again,
because the space such as we can learn about through our senses is
totally different from the space of the geometer's. Can we say
that geometry is based on the experience at all? Thorough
investigation will show us we can't. Therefore we may conclude
that these principles are in fact conditional assumptions\dots and
if we were transported into another world (which I call
non-Euclidean and try to describe), we would form different
assumptions. \dots Returning now to the question whether the
Euclidean geometry can be considered true or not, we must conclude
that it makes no sense. It would be the same as asking which
system is true --- metric one or that of old measures, or which
coordinates are more correct, Cartesian or polar ones. No geometry
is truer than the other; the question is only which is more
convenient for our purposes" (translated in English from
\cite[p.10]{3}).

\bigskip

This thesis of Poincar\'{e}'s concerning "convenience of a
specified geometry" has its merits. The principle of
geometrization expressed by all-encompassing formula (\ref{geom})
suggests there is a certain general geometry within the scope of
which any motion will be a free-fall. However, if such geometry
were founded, the question whether it would become a part of a
fundamental theory or remain a temporary half-empiric model
depends on its simplicity, and not in the least convenience.
Ptolemaic system can be seen as an example of a "working theory"\,
that can hardly be called simple or convenient. Poincar\'{e}'s
thesis can be applied to the historical relationship between
kinematics and geometry too. Indeed, the change of the kinematic
paradigm (i.e the switch from the geocentric to the heliocentric
model) drastically simplified description of the planet and star
motions. It is not a coincidence that only after the discoveries
of Copernicus, Brage and Kepler was Newton able to guess from the
simple elliptic trajectories of the planets the existence of one
of nature's simple and basic laws, the law of gravity. Further
analysis of this law and its consequences led to the geometric
paradigm being reconsidered (i.e. the switch from Euclidean
geometry to Riemannian one), and as a result all observable
disturbances of Kepler's ellipses can be now easily described
through characteristics of Schwarzschild's geodesic metrics in GR.

Our study helps to establish some of the limitations the
conventional approach of Poincar\'{e}'s has. If the relations
between geometry and kinematics were governed solely by
convenience, any more or less objective geometric world picture
would be impossible. But even through our over-simplified and
schematic approach one can glimpse obstacles standing in the way
of such absolute conventionalism. One such obstacle is the
existence of correlations (\ref{int7}) or (\ref{int8}) that limits
options for independent specification of the metric and kinematic
characteristics of the reference frame. The existence of this
correlation does not depend on concrete form of metric, i.e. it is
{\it metageometric} and therefore sets limits to Poincar\'{e}'s
conventionalism, the conventionality of the chosen geometry or
reference frame with its characteristics notwithstanding. Actually
the relations (\ref{int7})-(\ref{int8}) pose the only purely
mathematical obstacle that restricts the conventionalism discussed
above. In view of Poincar\'{e}'s scientific philosophy it is such
relations that can aspire to the role of "nature's laws", free
from convenience's influence.  Of particular interest is the
fundamental role which the rotation of the reference frame plays
in relations (\ref{int7})-(\ref{int8}).

\subsection{Complementarity of kinematics and geometry}

The aspect of relations (\ref{int7})-(\ref{int8}) mentioned
earlier is curiously analogous to the quantum mechanical
correlation between coordinate and momentum  uncertainty and the
quantum mechanical principle of complementarity. The uncertainties
relation in quantum mechanics restricts the degree of detalization
with which the state of quantum mechanical system  can be decribed
through coordinates and momentum. Mathematically the uncertainties
relation  is explained by non-commutativity of  dynamical
variables operators. Let us now turn back to GR and assume that
for some reason we need to explain disturbances of the law of
motion $\delta(t)$ in (\ref{z3}) and simultaneously correct the
metric (through $\delta g$) and reference frame (through
$\delta\overrightarrow{\tau}$). The relations
(\ref{int7})-(\ref{int8}) sets some fundamental limits to our
options here. In order to show more clearly their analogy with
quantum mechanics let us now represent (\ref{int7}) as follows:
\begin{equation}\label{intq}
\bar
g(\bar{\overrightarrow{a}},\bar{\overrightarrow{\omega}})+\tau(\Xi)=0,
\end{equation}
where $\bar{\overrightarrow{a}},\bar{\overrightarrow{\omega}}$ are
"physical vectors" of acceleration and angular velocity, i.e. they
consist of physical components of these vectors, $\bar g$ is
"physical metric" with components $\bar g_{\alpha\beta}\equiv
g_{\alpha\beta}/(H_\alpha H_\beta),$ $\Xi$ is 4-vector with
components
\begin{equation}\label{xiii}
\Xi=({}^{(3)}\text{div}\,\overrightarrow{\omega}/2,-D_t\overrightarrow{\omega}).
\end{equation}
After we've varied relation (\ref{intq}) over all values, we
obtain the equation\footnote{In detailed calculations it is
important to remember, that variation $\delta$ in general does'nt
commute with spatial partial derivatives $\partial_i$ in view of
relations (\ref{coordtr}), included in variational procedures.}:
\begin{equation}\label{varrr}
\delta\bar
g(\bar{\overrightarrow{a}},\bar{\overrightarrow{\omega}})+\bar
g(\delta\bar{\overrightarrow{a}},\bar{\overrightarrow{\omega}})+\bar
g(\bar{\overrightarrow{a}},\delta\bar{\overrightarrow{\omega}})+\delta\tau(\Xi)+\tau(\delta\Xi)=0.
\end{equation}
From  (\ref{varrr})  it follows that {\it there are fundamental
restrictions to the detalization of probe bodies motion in GR by
means of kinematics or geometry, and these restrictions are
expressed through the said integrability condition.} Here the
kinematic part of the variations $\delta \bar{\overrightarrow{a}},
\delta\bar{\overrightarrow{\omega}}$ can be seen as a relativistic
analogy of the kinematic values (coordinates) in the uncertainties
relation, whereas the metric part of variations $\delta \bar g$
can be seen as a relativistic analogy of dynamic values (momenta),
because metrics in GR satisfied  to the Einstein's dynamic
equations. Variation $\delta\tau$ of the reference frame does not
have a direct quantum mechanical analogy. Note that the
geometrical  anholonomicity of the space manifold measured by
$\omega$ serves as an analogy to quantum mechanical
non-commutativity of the variables. It is possible that
anholonomicity of manifolds is even more closely connected to
their quantization. Moreover, in this light the analogy between
quantum mechanics and gravity can be seen as much more than purely
superficial likeness. It gives us both more stronger motivation to
use geometric approach to quantization, and the appearance of the
new aspects of the gravity quantization problem including
geometry, kinematics and reference frames. Investigation of these
problems can be material for future studies.

\appendix

\section{Solution to the Killing equations for a general\\ spherically symmetric space-time}

For successful separation of variables in the system of equations
(1)-(10) in Sec. \ref{spher} it is important to find right order
of substitutions. With using (3) equations (8) and (9) takes the
following kind:
\begin{equation}\label{ee1}
{\tau^\theta}'=\frac{e^{2\lambda}}{r}\tau^\theta_{,\theta,\theta};\quad
{\tau^\varphi}'=\frac{e^{2\lambda}}{r\sin^2\theta}\tau^\theta_{,\theta,\varphi}.
\end{equation}
Differentiating (10) with respect to $r$ and substituting
(\ref{ee1}), we go to the third order equation for $\tau^\theta$:
\begin{equation}\label{ee2}
\sin^2\theta\left(\frac{\tau^\theta_{,\theta,\varphi}}{\sin^2\theta}\right)_{,\theta}+\tau^\theta_{,\theta,\theta,\varphi}=0.
\end{equation}
For the new variable  $X=\tau^\theta_{,\theta,\varphi}$ one can
obtain the following linear equation of the first order:
\[
X_{,\theta}-\cot\theta\, X=0,
\]
which has the solution
\begin{equation}\label{ee21}
X=\tau^\theta_{,\theta,\varphi}=\sin\theta\,\Phi_1(t,r,\varphi),
\end{equation}
where $\Phi_1$ is yet unknown function of the three mentioned
variables. Integrating this solution for
$\tau^\theta_{,\theta,\varphi}$ with respect to  $\theta$ and with
respect to  $\varphi,$ we go to the following general expression
for  $\tau^\theta$:
\begin{equation}\label{ee3}
\tau^\theta=-\cos\theta\Phi_1(t,r,\varphi)+\Phi_2(t,r,\varphi)+\Phi_3(t,r,\theta),
\end{equation}
where $\Phi_i$ are yet unknown function of the written
variables\footnote{We use economic notations if it is does'nt lead
to ambiguity. So, the functions  $\Phi_1$ in (\ref{ee21}) and  in
(\ref{ee2}) are different, but they depend on the same variables,
are  equally arbitrary and play the same role of multipliers in
front of the function, depending only on  $\theta.$ We may denote
the functions  in this situation by the same manner.}. Now taking
into account  (3), we obtain from  (\ref{ee2}):
\begin{equation}\label{ee4}
\tau^r=-r(\sin\theta\,\Phi_1(t,r,\varphi)+\Phi_{3,\theta}(t,r,\theta)).
\end{equation}
Differentiating (3) with respect to $\theta,$  expressing
$\tau^r_{,\theta}$ and substituting the result into  (8), we go to
the equation
\[
{\tau^{\theta}}'=\frac{e^{2\lambda}}{r}\tau^{\theta}_{,\theta,\theta},
\]
which after substitution (\ref{ee3}) gives the equation
\begin{equation}\label{ee5}
-\cos\theta\Phi_1'+\Phi_2'+\Phi_3'=\frac{e^{2\lambda}}{r}(\cos\theta\Phi_1+\Phi_{3,\theta,\theta}).
\end{equation}
Differentiating it with respect to  $\varphi$ and with respect to
$\theta,$ we go to the equation:
\[
\Phi_{1,\varphi}'+\frac{e^{2\lambda}}{r}\Phi_{1,\varphi}=0,
\]
whose integral has the form
\begin{equation}\label{ee6}
\Phi_1=\psi_1(t,\varphi)\mathcal{Q}+\psi_2(t,r),
\end{equation}
where
\begin{equation}\label{ee7}
\mathcal{Q}\equiv \exp\left\{-\frac{e^{2\lambda}}{r}\,
dr\right\};\quad e^{2\lambda}=-r\frac{\mathcal{Q}'}{\mathcal{Q}},
\end{equation}
and $\psi_2$ is functional constant of integration. Since it role
is to redefine  $\Phi_3,$ one can put $\psi_2$ equal to zero. Now
in view of equality of the first terms from the left and from the
right in (\ref{ee5}), we go to the following equations for
remainders:
\begin{equation}\label{ee8}
\Phi_2'+\Phi_3'=\frac{e^{2\lambda}}{r}\Phi_{3,\theta,\theta}.
\end{equation}
Differentiating (\ref{ee8}) with respect to $\varphi$ and taking
into account independence of $\Phi_3$ on $\varphi,$ we go to the
equation $\Phi_{2,\varphi}'=0,$ wherefrom it follows the
expression:
\begin{equation}\label{ee9}
\Phi_{2}(t,r,\varphi)=\phi_1(t,r)+\phi_2(t,\varphi).
\end{equation}
Coming back to the equation (\ref{ee8}), we obtain another
equation
\begin{equation}\label{ee10}
\frac{e^{2\lambda}}{r}\Phi_{3,\theta,\theta}-\Phi_3'=\phi_1'(t,r).
\end{equation}
Differentiating now (9) with respect to  $\varphi,$ and equation
(4) with respect to  $r$ and equating mixed derivatives
$\tau^{\varphi}_{,r,\varphi}=\tau^{\varphi}_{,\varphi,r},$ we go
to the following equation:
\[
-\frac{e^{2\lambda}}{r^2\sin^2\theta}\tau^{r}_{,\varphi,\varphi}=(\tau^\theta_{,\theta}-\tau^\theta\cot\theta)_{,r},
\]
or rewriting with using (\ref{ee3})-(\ref{ee4}), (\ref{ee6}),
(\ref{ee9}):
\begin{equation}\label{ee11}
\frac{\mathcal{Q}'\psi_{1,\varphi,\varphi}}{\sin\theta}=
\frac{\mathcal{Q}'\psi_1}{\sin\theta}+\Phi_{3,\theta}'-\cot\theta\Phi_{3}'-\cot\theta\,\phi'_1.
\end{equation}
Differentiating it with respect to  $\varphi$ and taking into
account independency of  $\Phi_3$ and $\phi_1$ on $\varphi,$ we go
to the following equation for $\psi_1$:
\[
\psi_{1,\varphi,\varphi,\varphi}+\psi_{1,\varphi}=0,
\]
whose general solution reads as follows:
\[
\psi_1=\psi_0(t)+C_1(t)\sin\varphi+C_2(t)\cos\varphi,
\]
where $\psi_0,C_1,C_2$ are yet unknown functional constant of
integration. Substituting this solution into (\ref{ee11}), we go
to the equation:
\[
\frac{\mathcal{Q}'\psi_0}{\sin\theta}+\Phi_{3,\theta}'-\cot\theta\Phi_{3}'-\cot\theta\,\phi'_1=0.
\]
Multiplying both parts of this equation by $\sin\theta$ and
differentiating the result on $\theta,$ we go to the following
equation:
\begin{equation}\label{ee12}
\Phi_{3,\theta,\theta}'+\Phi_3'+\phi_1'=0.
\end{equation}
Combining (\ref{ee12}) with (\ref{ee10}), we obtain
\[
\left(\Phi_3'+\frac{e^{2\lambda}}{r}\Phi_3\right)_{,\theta,\theta}=0.
\]
General solution to this equation has the following kind:
\begin{equation}\label{ee13}
\Phi_3(t,r,\theta)=\mathcal{Q}\chi_1(t,\theta)+\theta\chi_2(t,r)+\chi_3(t,r).
\end{equation}
Substituting it back into  (\ref{ee12}), we go to the expression
\begin{equation}\label{ee131}
\mathcal{Q}'\chi_{1,\theta,\theta}+\mathcal{Q}'\chi_1+\theta\chi_2'+\chi_3'+\phi_1'=0.
\end{equation}
Differentiating it twice with respect to $\theta$ we go to the
equation:
\[
\chi_{1,\theta,\theta,\theta,\theta}+\chi_{1,\theta,\theta}=0.
\]
Its general solution has the following kind:
\begin{equation}\label{ee14}
\chi_1=D_1(t)\sin\theta+D_2(t)\cos\theta+D_3(t)\theta+D_4(t),
\end{equation}
where $D_i(t)$ are yet unknown functional constants of
integration. Substituting this solution back into  (\ref{ee131})
and separating variables, we go to the following system:
\[
\mathcal{Q}'D_3+\chi_2'=0;\quad D_4\mathcal{Q}'+\chi_3'+\phi_1'=0.
\]
Integrating it with respect to $r$  and substituting all into
(\ref{ee13}), we obtain the following result:
\begin{equation}\label{ee15}
\Phi_3=\mathcal{Q}(D_1(t)\sin\theta+D_2(t)\cos\theta)+a(t)\theta+b(t),
\end{equation}
where $a,b$ are yet unknown functional constants of integration.
We have excluded from (\ref{ee15}) the term  $-\phi_1,$ since it
is cancelled with the term  $+\phi_1$ in (\ref{ee3}), which is
contained in  $\Phi_2$ by (\ref{ee9}). So, $\Phi_2$ becomes the
function of only two variables  $t$ and $\varphi.$ The equations
(\ref{ee10}) and (\ref{ee12}) are satisfied by  (\ref{ee15})
identically.

Now let us go to the integrability conditions for
$\tau^{\varphi}.$ Differentiating  (4) with respect to  $r,$  (9)
with respect to  $\varphi$ and equating mixed second derivatives
$\tau^\varphi_{,r,\varphi}=\tau^\varphi_{,\varphi,r},$ we go
(after some algebra) to the following equation:
\[
\psi_0(t)=D_2(t).
\]
It means that this values can be put to zero, since respective
terms in (\ref{ee3})-(\ref{ee4}) are cancelled. Differentiating
(4) with respect to  $\theta,$ (10) with respect to  $\varphi$ and
equating mixed second derivatives
$\tau^\varphi_{,\theta,\varphi}=\tau^\varphi_{,\varphi,\theta},$
we go to the expressions:
\[
a(t)=0;\quad \Phi_2=G_1(t)\sin\varphi+G_2(t)\cos\varphi-b(t).
\]
So, in expression for $\Phi_2$ the value  $-b$ can be omitted,
since  it is cancelled with  $b$ in (\ref{ee15}) for $\Phi_3$ in
all expressions. Now integrating (4), (9), (10) respectively over
$\varphi,$ $r$ and $\theta$ and equating results for
$\tau^\varphi,$ one can find general kind of this component. We
present general solution for spatial part of Killing equations in
the convenient for future purposes form:
\begin{equation}\label{taur}
\tau^r=-r\mathcal{Q}[\sin\theta\,
(C_1(t)\sin\varphi+C_2(t)\cos\varphi)+\cos\theta\,D(t)];
\end{equation}
\begin{equation}\label{tauth}
\tau^\theta=\mathcal{Q}[\sin\theta\, D(t)
-\cos\theta\,(C_1(t)\sin\varphi+C_2(t)\cos\varphi)]+G_1(t)\sin\varphi+G_2(t)\cos\varphi;
\end{equation}
\begin{equation}\label{tauph}
\tau^\varphi=-\frac{\mathcal{Q}}{\sin\theta}(C_1(t)\cos\varphi-C_2\sin\varphi)]+\cot\theta\,(G_1(t)\cos\varphi-G_2(t)\sin\varphi)+f(t),
\end{equation}
where all written functions are yet unknown.

Now let consider the equations  (5)-(7) and write their
integrability conditions with respect to the function $\tau^t,$
combining and equating different mixed second derivatives from
this component with using (\ref{taur})-(\ref{tauph}). The
integrability condition for $\tau^t$ with respect to  $\theta$ and
$\varphi$ leads to the expressions:
\begin{equation}\label{tp}
\dot G_i=0;\ (i=1,2)\quad \dot f=0.
\end{equation}
The integrability condition for $\tau^t$ with respect to  $\theta$
and $r$ adds the expression
\begin{equation}\label{tr}
e^{2\lambda-2\nu}r\dot{\widetilde{\mathcal{Q}}}_i=(r^2e^{-2\nu}\dot{\widetilde{\mathcal{Q}}}_i)_{,r},
\end{equation}
where $\widetilde{\mathcal{Q}}_i=C_i(t)\mathcal{Q},$ $i=1,2,$
$\widetilde{\mathcal{Q}}_3=D(t)\mathcal{Q}.$ The integrability
condition for $\tau^t$ with respect to  $r$ and $\varphi$ does'nt
add any new expressions. Now integrating the equations (5)-(7)
with respect to spatial variables and comparing the obtained three
expressions for $\tau^t$ we go to the following general expression
for  $\tau^t$:
\begin{equation}\label{taut}
\tau^t=-r^2e^{-2\nu}[(\dot{\widetilde{\mathcal{Q}}}_1\sin\varphi+\dot{\widetilde{\mathcal{Q}}}_2\cos\varphi)\sin\theta+\dot{\widetilde{\mathcal{Q}}}_3
\cos\theta]+\xi(t),
\end{equation}
where $\xi(t)$ is yet unknown function, together with one
additional to the (\ref{tr}) condition
\begin{equation}\label{cond0}
\dot{\mathcal{Q}}G_i=0,\ i=1,2.
\end{equation}

Easily to check, that  condition (\ref{tr}) has the following
integral:
\begin{equation}\label{intt1}
r^2e^{-2\nu}\dot{\widetilde{\mathcal{Q}}}_i\widetilde{\mathcal{Q}}_i=F_i(t),\
(i=1,2,3)
\end{equation}
where $F_i$ is arbitrary function of the time.

Substituting now expressions (\ref{taur})-(\ref{tauph}),
(\ref{taut}) into equations  (1)-(2) of the system of Killing
equations, we go to the following additional restrictions on the
metric, which must be satisfied for its integrability:
\begin{equation}\label{intt2}
(r\widetilde{\mathcal{Q}}_i)'+\dot\lambda
r^2e^{-2\nu}\dot{\widetilde{\mathcal{Q}}}_i+\lambda'r\widetilde{\mathcal{Q}}_i=0;
\end{equation}
\begin{equation}\label{intt3}
(e^{-2\nu}\dot{\widetilde{\mathcal{Q}}}_i)^{\cdot}+\dot\nu e^{-2\nu}
\dot{\widetilde{\mathcal{Q}}}_i+\frac{\nu'}{r}\widetilde{\mathcal{Q}}_i=0
\end{equation}
$i=1,2,3$ and the two more simple conditions:
\begin{equation}\label{intt4}
\xi(t)=\xi_0e^{-\nu};\quad \dot\lambda\xi_0=0;\quad
\xi_0=\text{const}.
\end{equation}

So, in general we have four conditions (\ref{tr}),
(\ref{intt2})-(\ref{intt4}) on the two metric functions  $\nu$ and
$\lambda,$ wherefrom it follows that integrability of the Killing
equations (1)-(10) takes place for only special cases. Using the
integral (\ref{intt1}), the equation (\ref{intt3}) can be reduced
to purely algebraic kind with respect to
$\widetilde{\mathcal{Q}}_i,$ which solutions reads as follows:
\begin{equation}\label{intt5}
\widetilde{\mathcal{Q}}_i=\sqrt{\frac{-\dot\nu F_i-\dot F_i\pm\sqrt{(\dot\nu F_i+\dot F_i)^2+4\nu'e^{2\nu}F_i^2/r}}{2\nu'r}}
\end{equation}
under $\nu'\neq0$ and
\begin{equation}\label{intt6}
\widetilde{\mathcal{Q}}_i=\pm\frac{F_ie^\nu}{r\sqrt{\dot F_i+\dot\nu F_i}}
\end{equation}
under $\nu'=0.$

Further analysis of compatibility of the system (1)-(10) is go
beyond the aims of our paper. We present some well known
particular isometry fields, satisfying all integrability
conditions:
\begin{enumerate}
\item the fields of rotation algebra of the group SO(3) ($C_i=0,$ $i=1,2,3,$ $\xi_0=0,$ $G_i\neq0,$ $f\neq0$):
\begin{equation}\label{sp3}
\tau_{(1)}=\cos\varphi\,\partial_\theta-\cot\theta\,\sin\varphi\,\partial_\varphi;\quad \tau_{(2)}=-\sin\varphi\,\partial_\theta-\cot\theta\,\cos\varphi\,\partial_\varphi;\quad
\tau_{(3)}=\partial_\varphi;
\end{equation}
\item for the metrics with $\dot\nu=0,$ $\dot\lambda=0$ ($C_i=0,$ $i=1,2,3,$ $\xi_0\neq0,$ $G_i=0,$ $f=0$) (static case):
\begin{equation}\label{sp4}
\tau_{(4)}=\partial_t.
\end{equation}
\end{enumerate}


\begin{thebibliography}{99}
\bibitem{1}
Yu. S. Vladimirov, {\it Reference frames in gravitation theory}\ \
M., Energoizdat, 1982 (In Russian).
\bibitem{2}
N. V. Mitskievich, {\it Relativistic physics in arbitrary
reference frame}\ \ arxiv: gr-qc/9606051.
\bibitem{sing}
L. G. Sing, {\it The theory of relativity,} M., Mir, 1963 (tr.
from English)
\bibitem{arno1}
V. I. Arnold, {\it Ordinary Differential Equations,} Izhevsk, UGU,
RCHD, 2000 (In Russian)
\bibitem{oneil}
B. O'Neill, {\it Semi-Riemannian geometry,} Akad. Press, Inc,
San-Diego, California, 1983.
\bibitem{3}
A. Poincar\'{e}, {\it Since and hypothesis,}\  In coll. of papers
"About Science"\ M., Nauka, 1990 (In Russian).
\bibitem{warner}
F. Warner, {\it Foundations of theory of smooth manifolds and Lie
groups,} Moscow, Mir, 1987 (tr. from English)
\bibitem{griffits}
F. Griffits, {\it External differential systems and calculus of
variations,} IO MFMI, 1999 (tr. from English)
\bibitem{rash}
P. K. Rashevski, {\it Uch. zapiski Mosk. Ped. Inst. im.
Libknehta,} (seriya fiz-mat.) N2, 83-94 (1938) (In Russian).
\bibitem{chow}
W. L. Chow, {\it Math. Ann.} {\bf 117}, N1 98-105 (1940).
\bibitem{koka}
S. S. Kokarev, {\it Introduction to General Relativity,}
Yaroslavl, YarSU, 2010.
\bibitem{koka2}
S. S. Kokarev, {\it Lectures "Elements of the theory of smooth
manifolds"\, (I): Lie derivatives and their applications,} In
Coll. papers of RSEC "Logos", Yaroslavl, v.4 (2009), pp. 77-166.
\bibitem{loshak}
G. Lochak,  {\it La geometrization de la physique,} Flammarion,
1994.
\bibitem{kok3}
S. S. Kokarev, {\it Three lectures on Newton's laws,} In Coll.
papers of RSEC "Logos", Yaroslavl, v.1, 2006, p.45-72, arXiv:
0905.3285v1[gr-qc]
\bibitem{odin}
S. Nojiri, S. D. Odintsov, {\it Phys. Rev.} {\bf D68}, 123512
(2003), hep-th/0307288
\bibitem{others}
S. S. Kokarev, {\it Gen.Rel.Grav,} {\bf 41}, pp. 1777-1794 (2009)

\end{thebibliography}
\end{document}